# Improper motions in lensed QSOs

L.L.R. Williams
Astronomy Department, PO Box 351580
University of Washington
Seattle, WA 98195-1580
llrw@astro.washington.edu

P. Saha
Mount Stromlo and Siding Spring Observatories
Australian National University
Canberra, ACT 0200, Australia
saha@mso.anu.edu.au

**Abstract:** We argue that individual images in multiple-image QSOs could easily have substructure at the level of $0.1''$ (i.e., unresolvable even with HST); microlensing within such substructure would cause centroid shifts, observable even from the ground as pseudo proper motions. We present a model of the four-image system 2237+0305 in which Image $B$ shows such "improper motions" of order of $0.01''$ over a few years.

## 1. INTRODUCTION

In multiple-image lensed QSOs the separate images are really not individual images but clumps of micro-images numbering of order the number of stars in the lensing galaxy. The standard approximation used to model such systems allows for graininess from individual stars but assumes no larger substructures within the galaxy. For typical surface densities the micro-images then cluster into macro-images of size $\sim 10^{-6}$ arcsec, with the macro-images separated by $\sim 1$ arcsec. The QSO optical continuum source size is estimated from microlensing as $\sim 10^{-7}$ arcsec—see e.g., Wambsganss *et al.,* (1990), Racine (1992).

Real galaxies, of course, have many substructures: spiral arms, globular clusters, and so on, not to mention any dark matter structures. Would these lead to image substructure on scales between $10^{-6}$ arcsec and 1 arcsec? Since HST images have shown no substructure within observed macro-images, any image substructure must be no larger than $\sim 0.1$ arcsec. Because of microlensing, unresolved structures can still have observable effects. Transverse motions of the observer, lens and source will cause independent brightness variations in different parts of any macro-image, causing the centroid of the macro-image to shift—a sort of pseudo proper motion which we will call improper motion. Since image centroids are routinely determined to $< 10^{-2}$ of the resolution, image substructure might be inferred even from the ground.[1] (High resolution radio observations would not help in this case because radio-emitting regions, being much larger than optical continuum sources, wash out microlensing.)

In this paper, we present a new model of 2237+0305 in which one of the images has substructure at the $\simeq 0.1$ arcsec level and consequent improper motions at the level of 0.01 arcsec in a few years. (Timescales for other multiple-image systems will be longer, but of the same order; we suggest how these could be estimated from our results by rescaling.) We started with one of several models in the literature and then mutated it by gradually redistributing some of the mass, while selecting for (a) better agreement with the macro-image observations, and (b) structure of extent $\simeq 0.1$ arcsec in one of the images. Having generated an extended image, we computed its microlensing properties. Our modeling method has some undesirable features resulting from the way it discretizes the mass distribution, but these blemishes are on the 0.01 arcsec scale for model macro-images, and do not carry over to the microlensing calculations. A more serious concern is that our model is highly non-unique, and we have not attempted to evaluate its probability relative to other models that may fit observations equally well without extended images. On the other hand, given how unremarkable the new model looks (Fig. 2), and how straightforward it was to concoct, the possibility of observable improper motions is not so easily dismissed now.

[1] The resolution of the naked eye is at best $\simeq 30$ arcsec, yet experienced observers can tell the difference between 0.7 arcsec and 0.5 arcsec seeing without a telescope (K.C. Freeman, personal communication).

## 2. A MACRO-MODEL

For this work we needed to find images of a point source for many different mass distributions, and compute changes in the images under small redistributions of the mass. Most importantly, we were concerned with splitting of the macro-images on the 0.1 arcsec scale. These considerations are rather different from those that previous numerical methods have been designed to meet. So we have developed a new approach.

The basic idea is to pixellate the image plane, including the mass distribution and the lens potential. Each pixel is a mass tile, and at the same time it is a square on the potential and time-delay surfaces, and also a square on the image plane. Pixels that are (relative to axial neighbors) maxima, minima, or saddles of the time delay are deemed to be image sites. The second derivatives that in the lensing formalism relate surface density, lens potential, and amplification are replaced by second differences. The pixel size ($s$ say—nothing to do with data pixels) is smaller than the macro-image substructure we will model, but much larger than micro-image separations; the results in this paper use $s = 0.01$ arcsec. We will refer to the images on pixels as mini-images, and each model macro-image will consist of one to several mini-images.

Concerning the second derivatives, $\partial^2/\partial x_1^2$ is replaced by $\Delta^2_{(11)}$, $\partial^2/\partial x_2^2$ by $\Delta^2_{(22)}$, and the mixed derivatives by $\Delta^2_{(12)}$ or $\Delta^2_{(21)}$, defined (for any pixellated function $f_{mn}$) as:

$$\begin{aligned} \Delta^2_{(11)} f_{mn} &= s^{-2}(f_{m+1,n} + f_{m-1,n} - 2f_{mn}) \\ \Delta^2_{(22)} f_{mn} &= s^{-2}(f_{m,n+1} + f_{m,n-1} - 2f_{mn}) \\ \Delta^2_{(12)} f_{mn} &= \Delta^2_{(21)} f_{mn} = \\ &\qquad (2s)^{-2}(f_{m-1,n-1} + f_{m+1,n+1} - f_{m+1,n-1} - f_{m-1,n+1}). \end{aligned} \tag{1}$$

The lens potential $\psi_{mn}$ is related to the surface density $\Sigma_{mn}$ by

$$\Sigma_{mn} \propto (\Delta^2_{(11)} + \Delta^2_{(22)})\psi_{mn}; \tag{2}$$

the proportionality factor does not matter for this work. With a point source at $(u, v)$—there is no advantage in discretizing the source position—the time delay is given by

$$\tau_{mn} = \tfrac{1}{2}(ms - u)^2 + \tfrac{1}{2}(ns - v)^2 - \psi_{mn}. \tag{3}$$

For any mini-image, the inverse amplification matrix is taken as

$$A^{-1} = \begin{pmatrix} 1 - \Delta^2_{(11)} & \Delta^2_{(12)} \\ \Delta^2_{(21)} & 1 - \Delta^2_{(22)} \end{pmatrix} \psi_{mn}, \tag{4}$$

and, as usual, the convergence $\sigma$ and shear $\gamma$ are given by

$$\begin{aligned} \mathrm{tr}(A^{-1}) &= 2(1 - \sigma) \\ \det(A^{-1}) &= (1 - \sigma)^2 - \gamma^2 \end{aligned} . \tag{5}$$

Discretizing the lens system thus makes it exactly soluble, but it has two uncomfortable side-effects. Firstly, there is no true magnification, since every image occupies exactly one pixel, and the amplification defined in (4) is artificial. Secondly, the system does not satisfy the odd-image theorem—in the continuous case non-singular lenses have first one extremum and then pairs of an extremum plus a saddle point in the time delay appear, whereas in the discrete system extrema can appear without saddles. To see this, suppose the $mn$-th pixel has value $(-1)^{m+n}$; then every pixel is an extremum and there are no saddles. If one now interpolates between the pixels, saddle points will appear near all corners between pixels. However, we can make the argument that the missing saddle points will have sharp curvature and contribute little flux. In the following, we will disregard these side-effects.

Changing any one of the $\psi_{mn}$ by some $\Delta\psi_{mn}$ amounts either to taking some of $\Sigma_{mn}$ and redistributing it evenly among its four axial neighbors, or the converse; the allowed range of any $\Delta\psi_{mn}$ is bracketed by the requirement that all $\Sigma_{mn} \geq 0$. This operation can cause mini-images to appear or disappear only at the $mn$-th pixel and its four axial neighbors, and $A$ can change only at these plus the four diagonal neighbors. Thus, one can redistribute the mass while keeping track of all mini-image changes by sweeping through the pixels, changing one $\psi_{mn}$ at a time, and comparing with eight neighbors. Needed now is a suitable prescription for choosing the $\Delta\psi_{mn}$. One such (and readers familiar with Ising magnets and their ilk will probably already have guessed it) is the following.

(i) Associate each mini-image with the closest observed image. Associated mini-images supply a flux, a centroid, and a size (i.e., spatial dispersion of the mini-images) for a model macro-image corresponding to each observed image.

(ii) Using the model fluxes, centroids and sizes, calculate a figure of merit $L$ for the agreement between model and observations. If the observations are well-enough understood, the likelihood is the appropriate $L$ to use; otherwise one just has to choose a sensible definition. Parity information could in principle be included, if available from orientations of radio jets; however 2237+0305 is radio quiet.

(iii) If a randomly chosen $\Delta\psi_{mn}$ would change $L$ by $\Delta L > 0$, accept that $\Delta\psi_{mn}$; if $\Delta L < 0$ accept that $\Delta\psi_{mn}$ with probability $1 - \Delta L/L$.

This is known as the Metropolis algorithm and in the many-iterations limit it will produce an ensemble of mass distributions weighted by $L$. It is discussed in many statistical mechanics books, for example Binney *et al.,* (1992).

To generate our new model of 2237+0305, we began with the potential

$$\psi = br + \tfrac{1}{2}\gamma r^2 \cos 2(\theta - \theta_\gamma) \tag{6}$$

with the parameter values given in Kochanek (1991). This is an isothermal sphere with critical radius $b$ and a relatively small quadrupole shear of strength $\gamma$ and orientation $\theta_\gamma$. We then pixellated within a circular window of radius 1.32 arcsec. The form of (6) makes the surface density completely circular in any window within which it applies. In the lensing galaxy the spiral arms are well outside of 1.32 arcsec, so assuming that the non-circular part of the mass is entirely outside this window is not a terrible approximation. We chose the source position to give the precise location for Image $A$, and kept it fixed throughout. Also fixed throughout were the mass distribution outside the 1.32 arcsec window, and the total mass—that is, we only move mass around within the window by varying the $\psi_{mn}$. We also stipulated that the convergence be subcritical ($\sigma < 1$) on all pixels except those on which the starting distribution (6) already had it supercritical (i.e., in the central $\simeq 0.5$ arcsec, well away from the images); this is a precaution against generating substructure with just a few super-massive pixels near the images. We plan to generalize in future work, but for now we attempt no estimates for the total mass or the improbability of the whole system. We defined $L$ on the basis of a fictitious set of observations. These were just the actual observations from Images $A$, $C$, and $D$, but we pretended that $B$ had been observed to have two components $B_1$ and $B_2$, 0.12 arcsec apart, with combined flux and centroid agreeing with the actual observations of $B$. Using this fictitious five-macro-image system, we took $L$ to be Gaussian in location, size, and relative flux, with the rather arbitrarily chosen dispersions of 0.025 arcsec in locations, 0.005 arcsec in size and 3–5% in fluxes. Then we proceeded with the Metropolis algorithm. After 375 sweeps over the window, the model macro-images $B_1, B_2$ were 0.1 arcsec apart, and fits in general had also improved (especially in flux). We then did a further 1000 sweeps, with the extra constraint that the locations of all maxima, minima, and saddles of the time delay be preserved—that is, we froze the mini-image locations but allowed the magnifications to vary. Finally, we averaged the potential over these last 1000 sweeps, recomputed the surface density and the mini-image locations and amplifications, and pooled the images associated with $B_1$ and $B_2$ into a single extended macro-image $B$ (thus discarding the fictitious observations and making any subsequent comparisons only with the actual observations). The result is our macro-model.

Figure 1 shows the fits of our model to the observed optical continuum positions and C III] fluxes. (It is advisable to use line fluxes for macro-models since the line-emitting region in QSOs is believed to be too large to be significantly affected by microlensing.) Simple parametrized models are usually very good at fitting image locations but rather poor at fitting fluxes (Kochanek [1991], Rix *et al.,* [1992], Wambsganss & Paczyński [1994] and references in these), whereas our pixellated model appears equally competent at both. This is, of course, what one expects—the image positions depend on the first derivatives of the potential, fluxes on the second derivatives, and a pixellated model provides freedom to make local adjustments to the potential while minimally altering its global properties, thus varying the flux with little effect on the image locations. The surface density of our model is in Fig. 2; it gives no hint that Image $B$ is extended.

Our model Image $B$ has three component mini-images, all of them extrema. Figure 3 shows the time-delay surface in the region of Image $B$. Their relative locations, and $\sigma$ and $\gamma$ values (listed in the caption) are inputs to the micro-model of the next section.

## 3. A MICRO-MODEL

The three mini-images that make up our model Image $B$ would themselves consist of micro-images. As the latter undergo flux variations due to stellar motions the centroid of the macro-image will shift. The amplitude and the potential observability of these shifts are the subject of this section.

To quantify this possible improper motion of the macro-images we used a ray tracing code as well as a simple model. A ray tracing code[2] based on a hierarchical tree code was used to generate 2D source plane amplification patterns for each of the three mini-images comprising Image $B$ in the last section. For each mini-image we calculate an amplification pattern frame $\sim 0.0001$ arcsec on the side, which corresponds to roughly 20 Einstein ring radii[3]. Each pixel corresponds to about $10^{-7}$ arcseconds. Each mini-image consists of a bundle of micro-images that are due to individual star-lenses (see Katz *et al.,* [1986] for an analytical treatment of micro-image distribution). Since the mini-images are much further apart than the frame size, each frame was calculated independently of

[2] with a couple of efficiency-enhancing modifications borrowed from Wambsganss (1990).

[3] Einstein ring radius of a lens of mass $M$ is given by $\theta_E = (\frac{4GM}{c^2}\frac{D_{\rm ls}}{D_{\rm ol}D_{\rm os}})^{1/2}$, where $D$'s are the angular diameter distances between observer, lens, and source.

the other two. The convergence and external shear parameters for each frame are those accompanying Fig. 3. In each frame all the lens mass was put into stars of Salpeter mass function, $dN(m)\,dm \propto m^{-2.35}\,dm$, with masses between 10 and 0.08 $M_\odot$. This is probably a reasonable assumption since the bulk of the mass close to galaxian centers is believed to be in stars with insignificant contributions from gas dark matter.

We assume that individual star positions are 'fixed' in the galaxy, but the entire galaxy has a proper motion of $600\,\mathrm{km\,sec}^{-1}$ in the lens plane. As the galaxy moves perpendicular to our line of sight the mini-images move across the corresponding critical lines in the lens plane and their amplifications and positions change with time. When a mini-image crosses a critical curve of a star-lens additional micro-images appear or disappear leading to a brightening or dimming of the whole mini-image. Now the macro-image centroid position can be calculated for a series of epochs, and also position *shifts* for pairs of epochs 1 and 10 years apart. Figure 4 shows the probability that the observed shift is greater than $\theta$ arcseconds if the $B$ image is observed 1, and 10 years apart. The least favorable probabilities are obtained if the two observations of the macro-images are made close together in time, say, 1 year apart, and the motion of the galaxy is perpendicular to shear. This is because the mini-images are elongated perpendicular to the shear and 1 year's worth of motion in the same direction barely moves the image a distance equal to its size.

Using a bulk velocity of the galaxy, as we have done here, is somewhat of an oversimplification. It has been shown (Kundić *et al.,* [1993]) that if the random velocity dispersion of stars in the lensing galaxy are considered instead of a bulk velocity of the galaxy the effective velocity of the caustic network becomes appreciably larger, by roughly 50%. Applied to our case this means that the actual time scale for observing QSO improper motions may be shorter by a third.

It is interesting to consider the limiting case of three mini-images where the optical depth is small, there is no shear, and the observations are well separated in time (i.e., $\gg R_{\rm lens}/v_{lens} \simeq 10\,\mathrm{yr}$ where $R_{lens}$ is the Einstein radius of a typical micro-lens and $v_{\rm lens}$ its velocity). The amplifications of the mini-images are then uncorrelated and are each given by the low optical depth probability distribution

$$p(A)\, dA = (A^2 - 1)^{-\frac{3}{2}}\, dA \tag{7}$$

(see, for example, Kaiser [1991]). Writing $z_i$ for the mini-image locations and assuming their separation $\gg R_{\rm lens}$, the centroid location is given by

$$\frac{\sum_i A_i z_i}{\sum_i A_i}, \tag{8}$$

where the $A_i$ are drawn independently from $p(A)$. The resulting cumulative centroid shift is plotted along with the ray tracing results as squares in Fig. 4. Even though this simple model underestimates the centroid shifts for observations made several years apart its general agreement with the more rigorous predictions is pleasing.

Even though the present paper looks at the particular case of 2237+0305, the general results can be easily extended to other QSOs whose image(s) are seen superimposed on a galaxy, or other similarly dense collection of stars. (For a treatment of microlensing in other quadruply-imaged QSO cases, see Witt, Mao & Schechter 1995). The timescales for detecting improper motions in other cases can be scaled from the ones derived in this paper. If the observer and the source are not moving in their respective planes (as was assumed in this paper) then the only relevant parameters are the angular velocity of the lens in the lens plane (here assumed to be 600 km/sec) and the distance to the lens. The source is assumed to be at typical QSO redshifts. The timescale is roughly proportional to the Einstein ring radius of the lens, and inversely proportional to the velocity of the lens in the lens plane. In effect,

$$t_{\rm improper} \sim \frac{\theta_{\rm microlens}}{v_{\rm lens}} \sim \frac{D_{\rm lens}^{-\frac{1}{2}}}{[D_{\rm lens}(1+z_{\rm lens})]^{-1}} = D_{\rm lens}^{\frac{1}{2}}(1+z_{\rm lens}). \tag{9}$$

Thus for typical lens redshifts the timescales for detecting improper motions, as measured by the observer, are not going to change appreciably compared to the case discussed in this paper.

## 4. DISCUSSION

The relative image positions for 2237+0305 found by Crane *et al.,* (1991) and by Rix *et al.,* (1992) differ at the 0.01 arcsec level; these are not significant, given their astrometric error estimates, though we cannot help being intrigued.

In our model the shift in the macro-image centroid position is mostly due to flux changes of the mini-images. Therefore one may expect the amplitude of flux changes to be positively correlated with the magnitude of the improper motion. In fact, for shifts less than half the image extent the mean flux change for a given centroid shift is well correlated with the magnitude of the shift, however the dispersion in the flux changes in large. This emphasizes that unlike proper motions, improper motions would not be steady drifts but comparatively sudden changes. A centroid shift of 10% of the image extent (the latter being chosen in our model to be $\simeq 0.1$ arcsec) corresponds to an average macro-image flux change of $\simeq 0.35$ mag. Images $A$ and $B$ both show flux changes of this order (Pen *et al.,* [1993], Racine [1992]) and our results suggest that these are accompanied by improper motions of order 10% of the image extent. If observed, improper motions would give an estimate of the extent of the macro-image.

Our model has a bearing on the attempt to get at the QSO optical continuum size, i.e., source size from micro-lensing lightcurves. Source size is most simply estimated as the ratio of the effective velocity of the source and the duration of a sudden observed magnitude change in macro-image lightcurve, see Racine (1992). If, however, the effective velocity of the source differs substantially from the assumed velocity then the source size estimate will be off by the corresponding factor. For a more rigorous determination of the source size one needs to know both the duration and the corresponding change in amplification of the image (Wambsganss *et al.,* [1990]). The latter gives you an independent leverage on the source size since smaller sources are more likely to get significantly amplified. Unfortunately, our model suggests that the observed sudden rise in amplification of, say image A, may be due to only one mini-image crossing a critical curve in the lens plane. In this case the observed amplification is only a lower limit on the actual amplification, the true amplification of the mini-image remains unknown, and so cannot be used to estimate the size of the source.

## ACKNOWLEDGEMENTS

We would like to thank Dr. Neal Katz for providing LLRW with the Barnes-Hut tree code and for many useful discussions of the code, Dr. Paul Schechter for urging PS to think about micro-images and the time delay surface, and the referee for several perceptive suggestions. LLRW acknowledges the support of the NASA grants NAGW 2569 and NAG 5 2793 at the University of Washington. PS is grateful to the University of Washington for the generous hospitality which made this collaboration possible.

## REFERENCES

Binney, J.J., Dowrick, N.J., Fisher, A.J., & Newman, M.E.J. 1992, *The theory of critical phenomena,* (Oxford Univ Press)

Crane, P., *et al.* 1991, ApJ., 369, L59

Kaiser, N. 1991. In *New Insights into the Universe*, p.247, eds. V. J. Martinez, M Portilla, & D. Saez, Proc. Summer School, Valencia, Spain

Katz, N, Balbus, S., & Paczyński, B. 1986, ApJ, 306, 2

Kochanek, C.S. 1991, ApJ, 373, 354

Kundić, T., Witt, H.J., & Chang, K. 1993. ApJ 409, 537

Pen U.-L., et al., 1993. In *Gravitational lenses in the Universe*, p.111, eds. J. Surdej *et al.,* Proc. 31st Liége Coll.

Racine, R. 1992, ApJ, 395, L65

Rix, H.-W., Schneider, D.P., & Bahcall, J.N. 1992, AJ, 104, 959

Wambsganss, J. 1990, Ph.D. Thesis

Wambsganss, J., Paczyński, B., & Schneider, P. 1990, ApJ, 358, L33

Wambsganss, J. & Paczyński, B. 1994, AJ, 108, 1156

Witt, H.-J., Mao, S., & Schechter, P.L. 1995, Preprint

## FIGURE CAPTIONS

**Figure 1.** Fits to the observed images: solid circles correspond to observations and dashed circles to our model. Centers of the solid circles correspond to positions from Rix et al., (1992), measured relative to Image $A$. Areas of the solid circles are proportional to C III] fluxes from Racine (1992). The filled circle near the center shows the model source position. The numerical values of positions and relative fluxes are as follows.

| $\Delta\alpha_{\rm obs}$ | $\Delta\alpha_{\rm mod}$ | $\Delta\delta_{\rm obs}$ | $\Delta\delta_{\rm mod}$ | $\rm flux_{obs}$ | $\rm flux_{mod}$ |
|---|---|---|---|---|---|
| 0.000 | 0.000 | 0.000 | 0.000 | 0.312 | 0.273 |
| -0.676 | -0.673 | 1.686 | 1.680 | 0.319 | 0.278 |
| 0.625 | 0.620 | 1.200 | 1.202 | 0.185 | 0.252 |
| -0.869 | -0.860 | 0.520 | 0.540 | 0.184 | 0.196 |

**Figure 2.** Surface density in our model of the central $2 \times 1.32$ arcsec of the lens, with the model macro-image positions and relative fluxes again. If the mass distribution had constant mass-to-light then consecutive contours would differ by half a magnitude. The contour labelled 'crit' corresponds to the critical surface density ($\sigma = 1$), which for 2237+0305 is $3.36h^{-1}\,{\rm gm\,cm^{-2}}$ or $4.65 \times 10^9 h^{-3} M_\odot\,{\rm arcsec^{-2}}$.

**Figure 3.** Time delay contours near Image $B$ in our model, with dashed circles indicating positions (positions) and relative fluxes (areas) of the mini-images. Note, however, that our model is completely discrete with a pixel size of 0.01 arcsec; for this plot we have interpolated between the pixels, so it should not be taken too literally—in particular, interpolation shifts the image positions by up to a pixel size. The parameters of the mini-images making up our model Image $B$ are as follows.

| $\Delta\alpha$ | $\Delta\delta$ | $\sigma$ | $\gamma$ | $A$ |
|---|---|---|---|---|
| $-0.65$ | 1.70 | 0.422 | 0.446 | 7.40 |
| $-0.69$ | 1.67 | 0.194 | 0.311 | 1.81 |
| $-0.73$ | 1.63 | 0.246 | 0.408 | 2.49 |

**Figure 4.** Probability of observing a displacement in the macro-image centroid position of more than $\theta$ arcseconds in 1, and 10 years if the lensing galaxy has a bulk proper motion of $600\,\mathrm{km\,sec^{-1}}$. The lines are obtained from the microlensing model: dashed for proper motion along the shear and solid for proper motion perpendicular to the shear. The squares are predictions of the simple model at the end of Section 3.

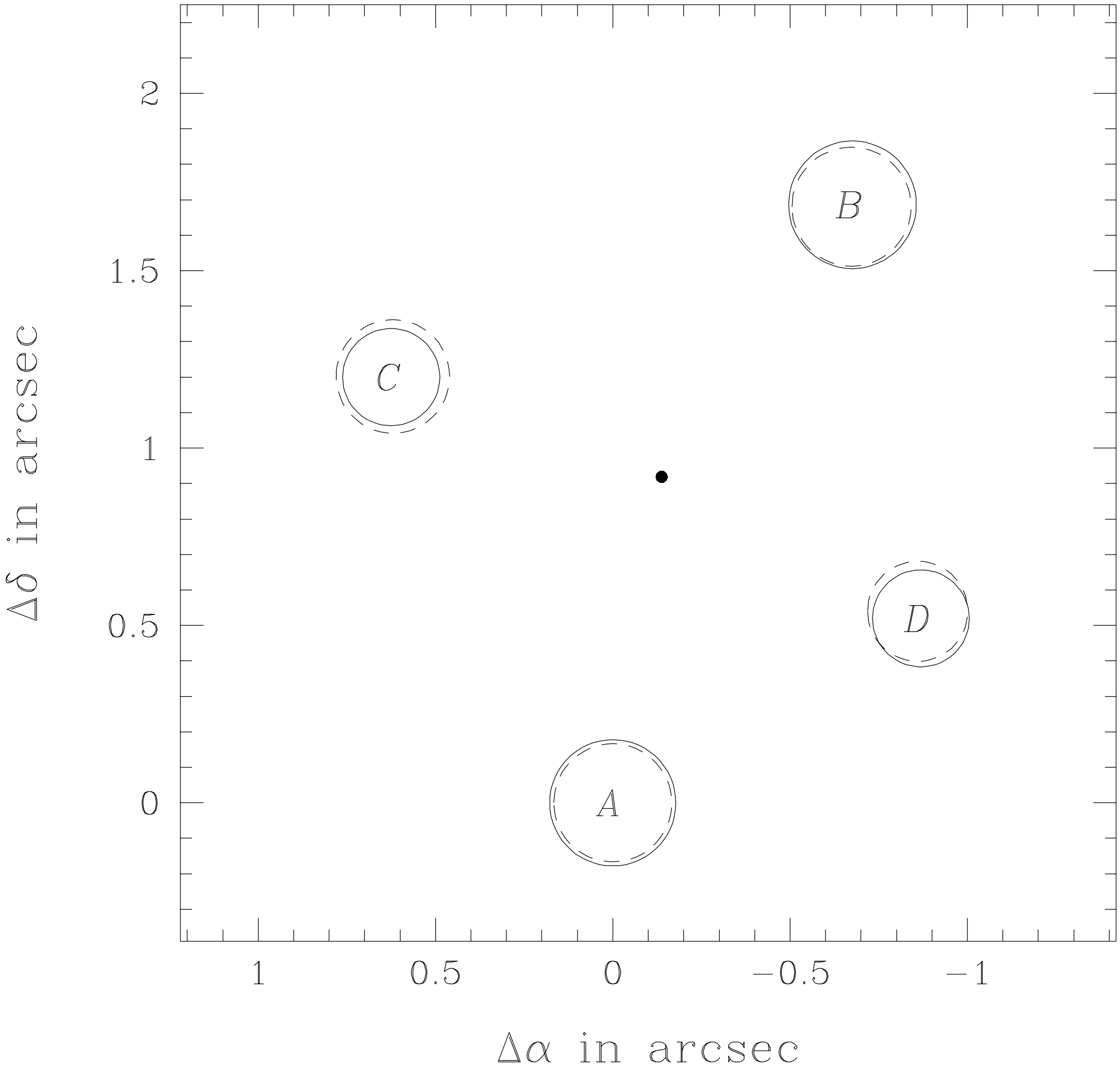

**Figure 1**

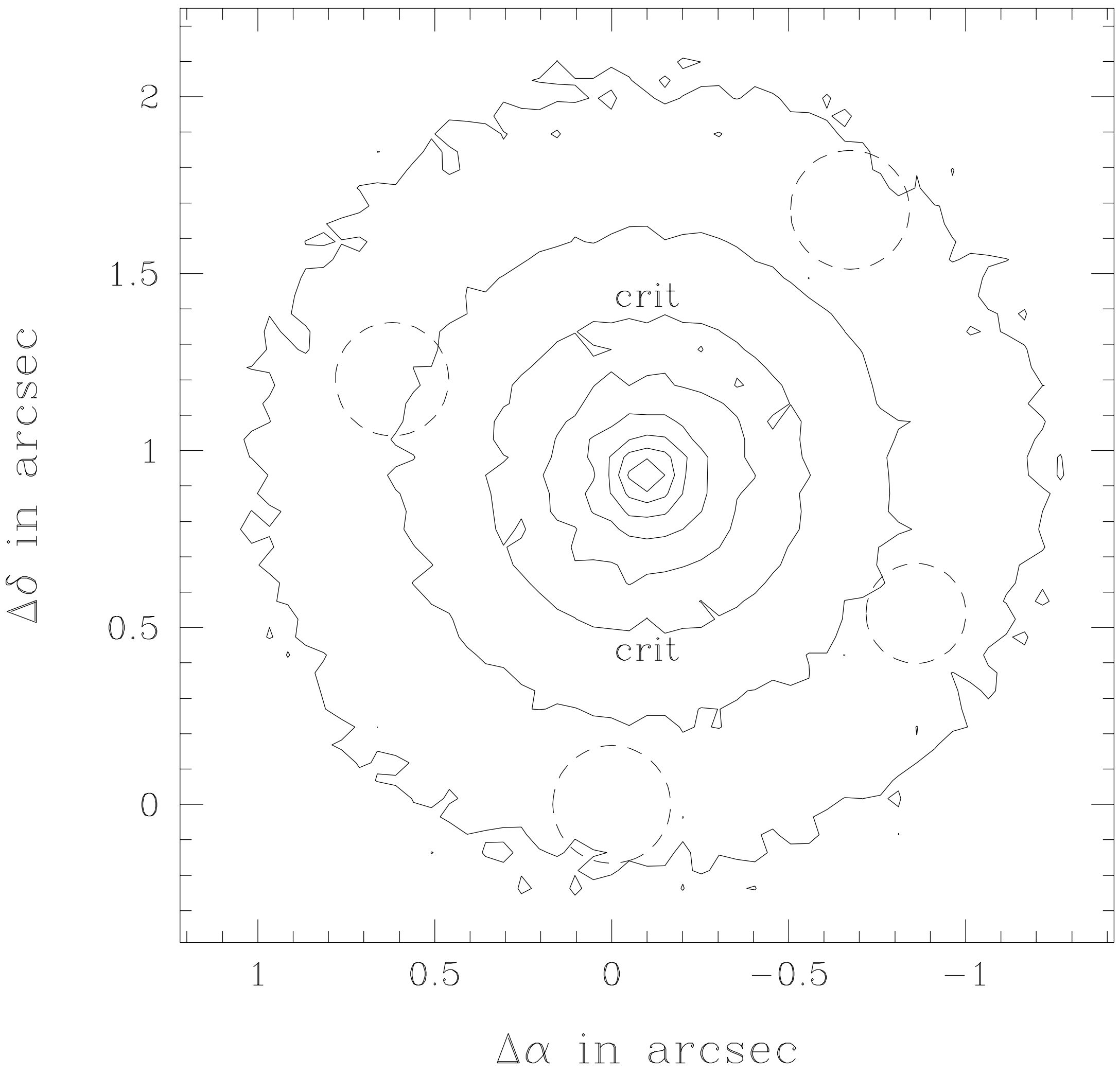

**Figure 2**

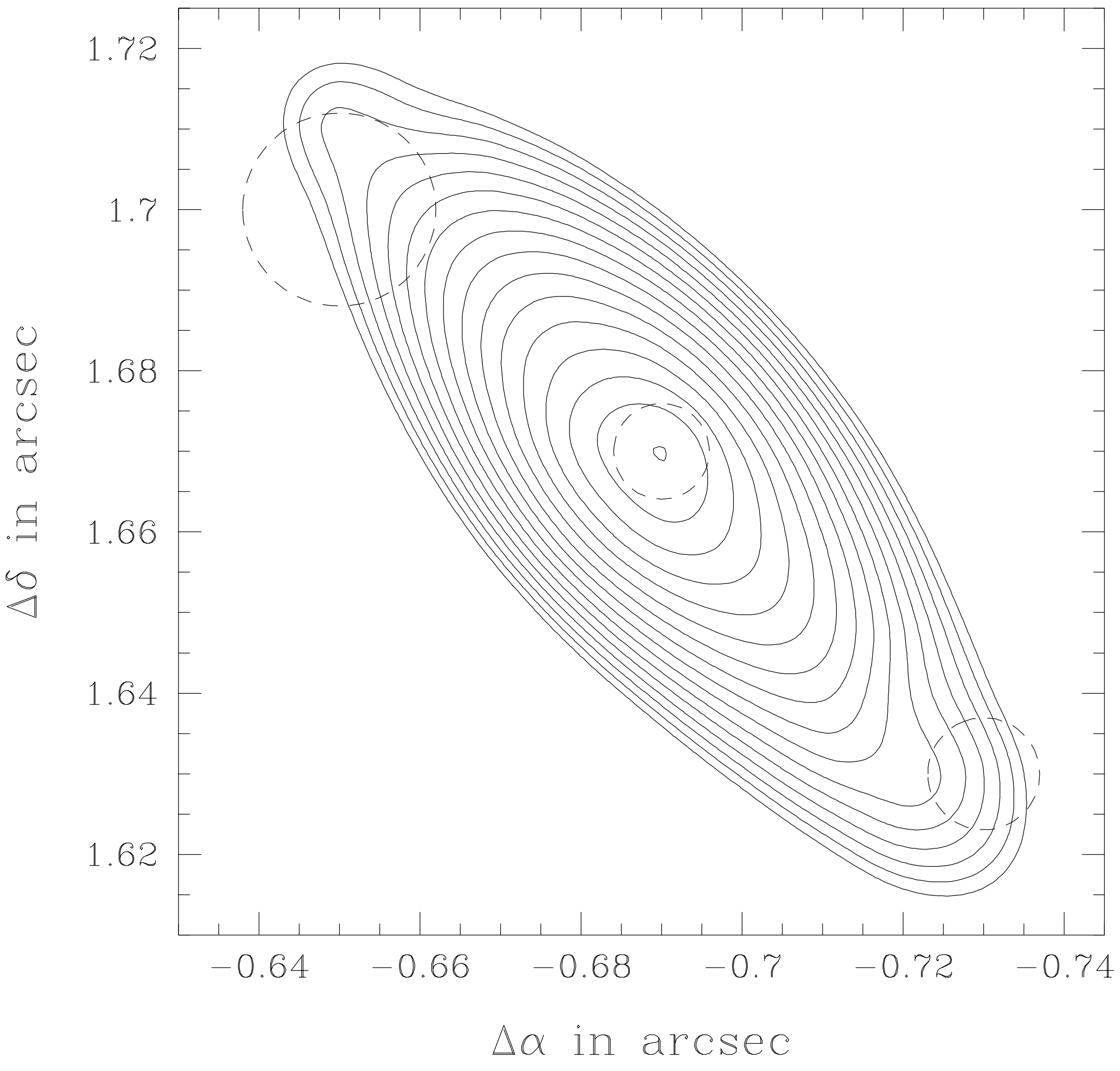

**Figure 3**

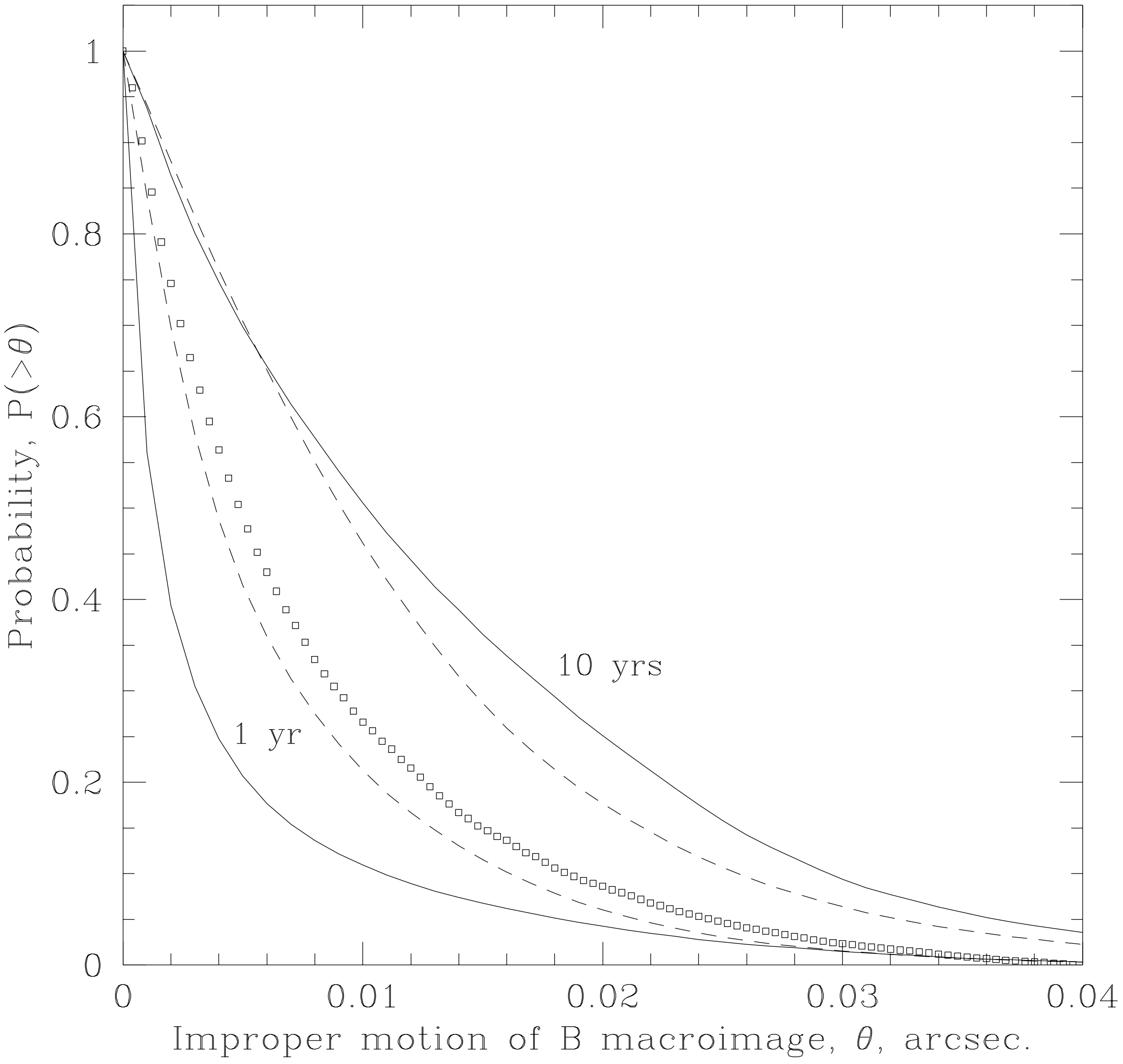

**Figure 4**